\documentclass[aps,pre,twocolumn,floatfix,showpacs,showkeys]{revtex4-1}

\usepackage{graphicx}
\usepackage{natbib}
\usepackage{epstopdf}
\usepackage{amsmath}
\usepackage{amssymb}
\usepackage{mathbbol}
\usepackage{braket}
\usepackage[hyperindex,colorlinks,urlcolor=blue]{hyperref}

\usepackage{xcolor}

\usepackage{tikz}
\usepackage{subfigure}

\begin{document}

\title{Spontaneous emission from radiative chiral nematic liquid crystals at the photonic band gap edge: an investigation into the role of the density of photon states near resonance} 

\author{Th. K. Mavrogordatos}\email{ttm22@cam.ac.uk}
\author{S. M. Morris}\email{smm56@cam.ac.uk}
\author{S. M. Wood}\email{smw73@cam.ac.uk}
\author{H. J. Coles}\email{hjc37@cam.ac.uk}
\author{T. D. Wilkinson}\email{tdw13@cam.ac.uk}
\affiliation{Centre of Molecular Materials for Photonics and Electronics, Department of Engineering, University of Cambridge,
9 JJ Thomson Avenue, Cambridge CB3 0FA, United Kingdom}

\begin{abstract}
In this article, we investigate the spontaneous emission properties of radiating molecules embedded in a chiral nematic liquid crystal, under the assumption that the electronic transition frequency is close to the photonic edge mode of the structure, i.e., at resonance. We take into account the transition broadening and the decay of electromagnetic field modes supported by the so-called `mirror-less' cavity.  We employ the Jaynes-Cummings Hamiltonian to describe the electron interaction with the electromagnetic field, focusing on the mode with the diffracting polarization in the chiral nematic layer. As known in these structures, the density of photon states, calculated via the Wigner method, has distinct peaks on either side of the photonic band gap, which manifests itself as a considerable modification of the emission spectrum. We demonstrate that, near resonance, there are notable differences between the behavior of the density of states and the spontaneous emission profile of these structures. In addition, we examine in some detail the case of the logarithmic peak exhibited in the density of states in 2D photonic structures and obtain analytic relations for the Lamb shift and the broadening of the atomic transition in the emission spectrum. The dynamical behavior of the atom-field system is described by a system of two first order differential equations, solved using the Green's function method and the Fourier transform. The emission spectra are then calculated and compared with experimental data.
\end{abstract}

\keywords{chiral nematic liquid crystals, photonic density of states, spontaneous emission, Lamb shift}
\pacs{42.70.Df, 42.70.Qs, 42.50.−p, 78.15.+e}

\date{\today}

\maketitle

\section{Introduction}
\label{sec:intro}

The concept of the density of photon states (DOS) is regularly employed to study the photonic crystal properties that determine emission and absorption of electromagnetic radiation of a given frequency from guest atoms \cite{DOSPC}. For the case of (dye-doped) chiral nematic liquid crystal (LC) films, which constitute a representative example of partial 1D photonic crystals, the link between the behavior of the DOS and fluorescence is attempted in \cite{DOSfl}, where Dirac's rule (often referred to as `Fermi's golden rule') is employed to calculate the photon emission rate. With regard to the interaction of a gain medium and the electromagnetic field in the resonant cavity, a two-level system coupled to a quantum harmonic oscillator is frequently described with the Jaynes-Cummings (JC) Hamiltonian in which only `resonant' terms feature \cite{RWAIrish, HarocheBook, PerlinDOS}. Such a consideration is permissible in the case of near resonance and weak coupling \cite{RWAIrish}. Both conditions are satisfied for spontaneous emission in these periodic structures for small detuning \cite{HarocheBook}. It should be mentioned here, that through a change of basis, a much greater range of coupling strength and detuning values can be allowed \cite{RWAIrish}. In \cite{PerlinDOS} the JC model is employed to describe emission in the vicinity of the saddle point of 2D photonic crystals, where the DOS exhibits a logarithmic peak. In order to analyze the behavior of the DOS near its logarithmic peak and at the edge of the band gap, one can resort to a critical point analysis through an expansion in the region of the saddle point \cite{DOSPC}.

In this article, we put the analysis of spontaneous emission from 2D photonic crystals on a firmer basis providing analytical results, and explore in more detail the fluorescence properties in chiral nematic LCs, outlining common features that are attributed to resonance. Moreover, the discrepancy between the experimentally obtained emission spectra and the theoretically calculated DOS is addressed. Such a consideration aims to further the understanding of spontaneous and induced emission from these distributed feedback resonators. The case of chiral nematic LCs is selected because these structures have the additional advantage of allowing an exact analytic solution of Maxwell's equations \cite{BelyakovOEM}.

\section{The Jaynes-Cummings model in a resonant environment}
\label{sec:JCresenv}

We will now outline the basic formulation of an atom-field interaction in a resonant environment, within the framework of the JC model. In many cases, the frequency of the electric field in a resonant structure is near the transition frequency of a two-level system. Such a two-level system is called an `atom' for convenience \cite{BykovBook}. The JC Hamiltonian for the atom-field system is written in the form \cite{RWAIrish, BykovBook, PerlinDOS}:
\begin{align}\label{JCeq}
H=\frac{1}{2}\hbar\omega_{10}\sigma_z+\hbar\sum_{\kappa}\omega_{\kappa}a^{\dagger}_{\kappa}a_{\kappa}+
i\sum_{\kappa} (\mu_{\kappa}a^{\dagger}_{\kappa}\sigma_{-}-\mu_{\kappa}^{*}a_{\kappa}\sigma_{+}),
\end{align}
where:
\begin{align*}
  \mu_{\kappa}=(\mathbf{d}\cdot\hat{\mathbf{e}}_{kl})\omega_{10}\sqrt{\frac{2\pi\hbar}{V\varepsilon\omega_{\kappa}}}\:,
\end{align*}
is the atom-field coupling constant, $\omega_{10}$ is the atomic transition frequency, $\omega_{\kappa}$ is the electromagnetic mode frequency,  $\sigma_z$ is the inversion operator, $\sigma_{\pm}={(\sigma_x\pm i \sigma_y)}/{2}$ are the raising and lowering operators for Pauli matrices acting on qubit states, $a^{\dagger}_{\kappa}$ and $ a_{\kappa}$ are the bosonic creation and annihilation operators for the $\kappa^{th}$ mode,  $\mathbf{d}=e\mathbf{r}_{10}$ is the transition dipole moment, $\hat{\mathbf{e}}_{kl}$ is the unit field polarization vector and $\varepsilon$ is the frequency independent dielectric constant of the medium with volume $V$. In writing the  Hamiltonian in the form of Eq. \eqref{JCeq}, we assume that the energies of the upper and lower states of the atom are equal and opposite, i.e., $E_1=-E_0=(1/2)\hbar\omega_{10}$.

The first term in the interaction part of the JC Hamiltonian in the so-called `rotating field' approximation corresponds to the electronic transition from the lower to the upper level with the absorption of a photon from the field mode, while the second term represents the reverse process, i.e., the electronic transition from the upper to the lower atomic level and the emission of a photon that contributes into the mode field. The remaining terms in the interaction Hamiltonian [of the form $H_{\rm int}=G(\sigma_{+}+\sigma_{-})(E^*a^{\dagger}+Ea)$] that do not feature in Eq. \eqref{JCeq} correspond to non-resonant, virtual processes \cite{BykovBook}, namely to the electronic transition upwards accompanied by the emission of one photon and the transition downward accompanied by the absorption of one photon. Due to their smallness, these terms are usually omitted \cite{PerlinDOS, BykovBook}. We consider the atom-field system to be described by the wavefunction:
\begin{align}\label{wfunc}
  &\Ket\psi=\exp(-i\omega_{10}t/2)\big[c_1(\mathbf{R},t)\Ket{1,\{0\}}\\
  &+\sum_{\kappa}\exp(i\delta_{\kappa}t)c^{\kappa}_{0}(\mathbf{R},t)\Ket{0,\{1\}_{\kappa}}\big]\;,
\end{align}
where: $\delta_{\kappa}=\omega_{10}-\omega_{\kappa}$. This expression is a superposition of the state $\Ket{1,\{0\}}$, corresponding to the excited state of the atom with the photon occupation number equal to 0, and the state $\Ket{0,\{1\}_{\kappa}}$, corresponding to the ground state with one photon in the $\kappa^{\rm th}$ mode. Hence, the mechanism of spontaneous emission is quantified. In order to account for cavity losses and linewidth broadening, we introduce phenomenologically a small imaginary part to the field and transition frequencies, respectively, such that \cite{PerlinDOS}:
\begin{align*}
\omega_{\kappa}=\acute{\omega_{\kappa}}-i\gamma \: \text{ and } \:\omega_{10}=\acute{\omega_{10}}-i\gamma_{10}.
\end{align*}
Expressions for the imaginary part of the edge mode frequencies (eigenfrequencies) in the case of a chiral nematic LC layer can be found in \cite{BelyakovOEM}. The appearance of such an imaginary contribution is related to the energy leakage through the surfaces of the LC film. Substituting the wavefunction of Eq. \eqref{wfunc} into the time-dependent Schr\"{o}dinger equation yields a system of the equations of motion [see Appendix \ref{sec:AppA}], which is reduced to:
\begin{align}\label{dercoeff}
  \frac{dc_1}{dt}(\mathbf{R},t)=-\int_0^t g(\mathbf{R},t-\tau)c_{1}(\mathbf{R},\tau)d\tau\;,
\end{align}
with the Green's function taking the form of:
 \begin{align*}
 g(\mathbf{R},t)&=u(t)\sum_{\kappa}|\mu_{\kappa}|^2\exp(i\delta_{\kappa}t)\equiv\beta u(t)\\
 &\times \int_{\omega_1}^{\omega_2}\frac{\exp{[i(\omega_{10}-\omega) t]}}{\omega-i\gamma}\rho_{l}(\mathbf{R},\omega)d\omega\;,
 \end{align*} 
in which $\rho_l(\mathbf{R},\omega)$ is the local density of photon states, $u(t)$ is the unit step function and:
 \begin{align*}
   \beta\propto\frac{\omega_{10}^2|\mathbf{r}_{10}|^2}{\varepsilon S},
 \end{align*}
where $S$ is the area of the LC layer.

The emission spectrum is then determined by the relation [see Appendix \ref{sec:AppB}]:
\begin{equation}\label{spectrum}
\begin{aligned}
&S(\Omega)=2\Re\{\tilde{c_1}(\Omega-\omega_{10})\}\\
&=\frac{2[\gamma_{10}+\Sigma_2(\Omega)]}{[\Omega-\acute{\omega_{10}}+\Sigma_1(\Omega)]^2+[\gamma_{10}+\Sigma_2(\Omega)]^2}\;,
\end{aligned}
\end{equation}
with:
\begin{align*}
\tilde{c_1}(\Omega-\omega_{10})=\int_0^{\infty}c_1(t)e^{i(\Omega-\omega_{10}) t}dt
\end{align*}
and \cite{PerlinDOS}:
\begin{equation}\label{S12}
\begin{aligned}
&\Sigma_1={\beta}|{\hat{\mathbf{e}}_{kl}(\mathbf{R})\hat{\mathbf{r}}_{10}}|^2\Re\left\{\int_{\omega_{\rm min}}^{\omega_{\rm max}}\frac{\rho(\omega)}{(\omega-i\gamma)(\omega-\Omega-i\gamma)}d\omega\right\},\\
&\Sigma_2={\beta}|{\hat{\mathbf{e}}_{kl}(\mathbf{R})\hat{\mathbf{r}}_{10}}|^2\Im\left\{\int_{\omega_{\rm min}}^{\omega_{\rm max}}\frac{\rho(\omega)}{(\omega-i\gamma)(\omega-\Omega-i\gamma)}d\omega\right\},
\end{aligned}
\end{equation}
where $\rho(\omega)$ is the density of photon states (DOS). The term $\Sigma_1$ is linked to the Lamb shift, while the term $\Sigma_2$ is related to the broadening of the transition between the two states of the system.

We will next consider the characteristic case of chiral nematic LCs, which are partial 1D photonic crystals, where the DOS can be derived from the transmission properties of a layer with finite thickness \cite{DOSfl}, enabling the prediction of the emission spectrum using Eqs. \eqref{spectrum} and \eqref{S12}.

\section{The density of photon states for chiral nematic liquid crystals}
\label{sec:DOSCLC}

For the extraction of the transmission coefficient and subsequently the density of photon states in chiral nematic LCs, we proceed as follows. We consider a boundary value problem formulated such that two plane waves of the diffracting polarization are incident on a chiral nematic LC layer. The assumption of no boundary reflection allows the separation of eigenpolarizations, introducing an error of the order of the relative dielectric anisotropy \cite{BelyakovBook}. By demanding a continuous tangential component of the electric and magnetic field at the layer-glass interface, we formulate a system, the solution of which yields the transmission coefficient for light of diffractive circular polarization. The transmission coefficient for a layer with $N$ full precessions of the molecular director, and hence thickness $L=Np$, reads \cite{BelyakovSemenov2}:
\begin{equation}\label{Tcoeff}
T=\frac{\exp{\left(\frac{i\tau L}{2}\right)}\left(\dfrac{q\tau}{k^2}\right)}{\dfrac{q\tau}{k^2}\cos(qL)+i\left[\left(\dfrac{\tau}{2k}\right)^2+\left(\dfrac{q}{k}\right)^2-1\right]\sin(qL)}\;,
\end{equation}
where
\begin{equation*}
q=k\sqrt{1+\left(\frac{\tau}{2k}\right)^2-\sqrt{\left(\frac{\tau}{k}\right)^2+{\delta}^2}}.
\end{equation*}
In these expressions, $\tau={(4\pi)}/{p}$, $k={(\omega}/{c})\epsilon_0$, with $\epsilon_0={(\epsilon_{\parallel}+\epsilon_{\perp})}/{2}$ being the average dielectric constant where $\epsilon_{\parallel}$ and $\epsilon_{\perp}$ are the relative dielectric constants parallel and perpendicular to the director, respectively, and $\delta={(\epsilon_{\parallel}-\epsilon_{\perp})}/{(\epsilon_{\parallel}+\epsilon_{\perp})}$ the relative dielectric anisotropy, $p$ the helical pitch and $c$ the speed of light in the vacuum. This relation is a different expression to the one given in \cite{DOSfl}, where Maxwell's equations are solved in a frame rotating with the molecular director. Omitting common real prefactors and frequency independent terms, the real and imaginary parts, respectively, of the transmission coefficient [Eq. \eqref{Tcoeff}] read:
\begin{align}\label{XY}
&X=\left(\frac{q\tau}{k^2}\right)\cos(qL),\\
&Y=-\left[\left(\frac{\tau}{2k}\right)^2+\left(\frac{q}{k}\right)^2-1\right]\sin(qL).
\end{align}
The normalized DOS can be written as \cite{DOSfl}:
\begin{equation}\label{normDOS}
\rho=\frac{c}{Np\sqrt{\epsilon_0}}\frac{X\dfrac{dY}{d\omega}-Y\dfrac{dX}{d\omega}}{X^{2}+Y^{2}}.
\end{equation}
Focusing in the region of the band gap, we can write: $q=i\tilde{q}$. The real and imaginary parts become, respectively,
 \begin{align*}
\end{align*}
One can easily verify that in this region:
\begin{equation*}
\tilde{\rho}=\frac{c}{Np\sqrt{\epsilon_0}}\frac{\tilde{X}\dfrac{d\tilde{Y}}{d\omega}-\tilde{Y}\dfrac{d\tilde{X}}{d\omega}}{\tilde{X}^{2}+\tilde{Y}^{2}}=\frac{c}{Np\sqrt{\epsilon_0}}\frac{X\dfrac{dY}{d\omega}-Y\dfrac{dX}{d\omega}}{X^{2}+Y^{2}}=\rho.
\end{equation*}
Hence, we find that the expressions in Eqs. \eqref{XY} and \eqref{normDOS} can also be used inside the band gap. The low frequency band-edge for a chiral nematic LC is given by the formula $\omega_0={\omega_c}/{\sqrt{1+\delta}}$ where $\omega_c={(2\pi c)}/{p}$ is the center of the reflection band \cite{BelyakovOEM}. In our treatment, we approximate the edge mode frequency in the long wavelength edge with $\omega_0$. For a layer with finite thickness, this approximation increases in validity as $N\delta$ is appreciably higher than unity \cite{BelyakovOEM}. The expressions in Eqs. \eqref{S12} should be averaged over all possible orientations of the dipole moment and from this procedure we obtain the pertinent transition dipole order parameter, as shown in \cite{DOSfl}. In what follows, we will assume that the dipole order parameter is zero; this corresponds to an isotropic distribution of the dyes i.e., absence of preferential alignment. We also consider a uniform distribution of the fluorescent molecules in the chiral nematic host, so that our results are not affected by the spatial distribution of the eigenmodes. In \cite{DOSfl} it is shown that using these assumptions $\big<|{\hat{\mathbf{e}}_{kl}(\mathbf{R})\hat{\mathbf{r}}_{10}}|^2\big>=1/3$.
\begin{figure}
\includegraphics[width=2.8in]{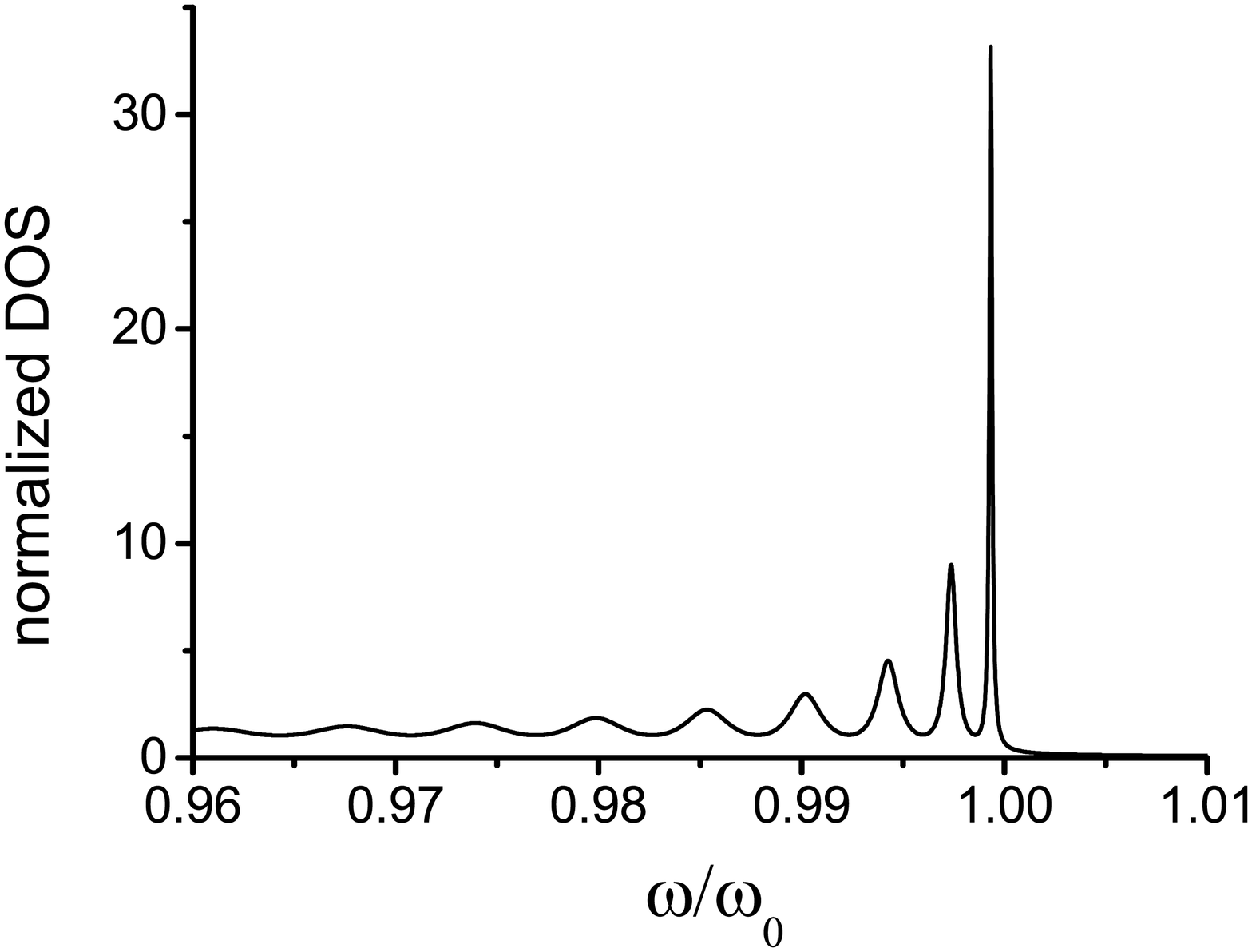}
\includegraphics[width=2.8in]{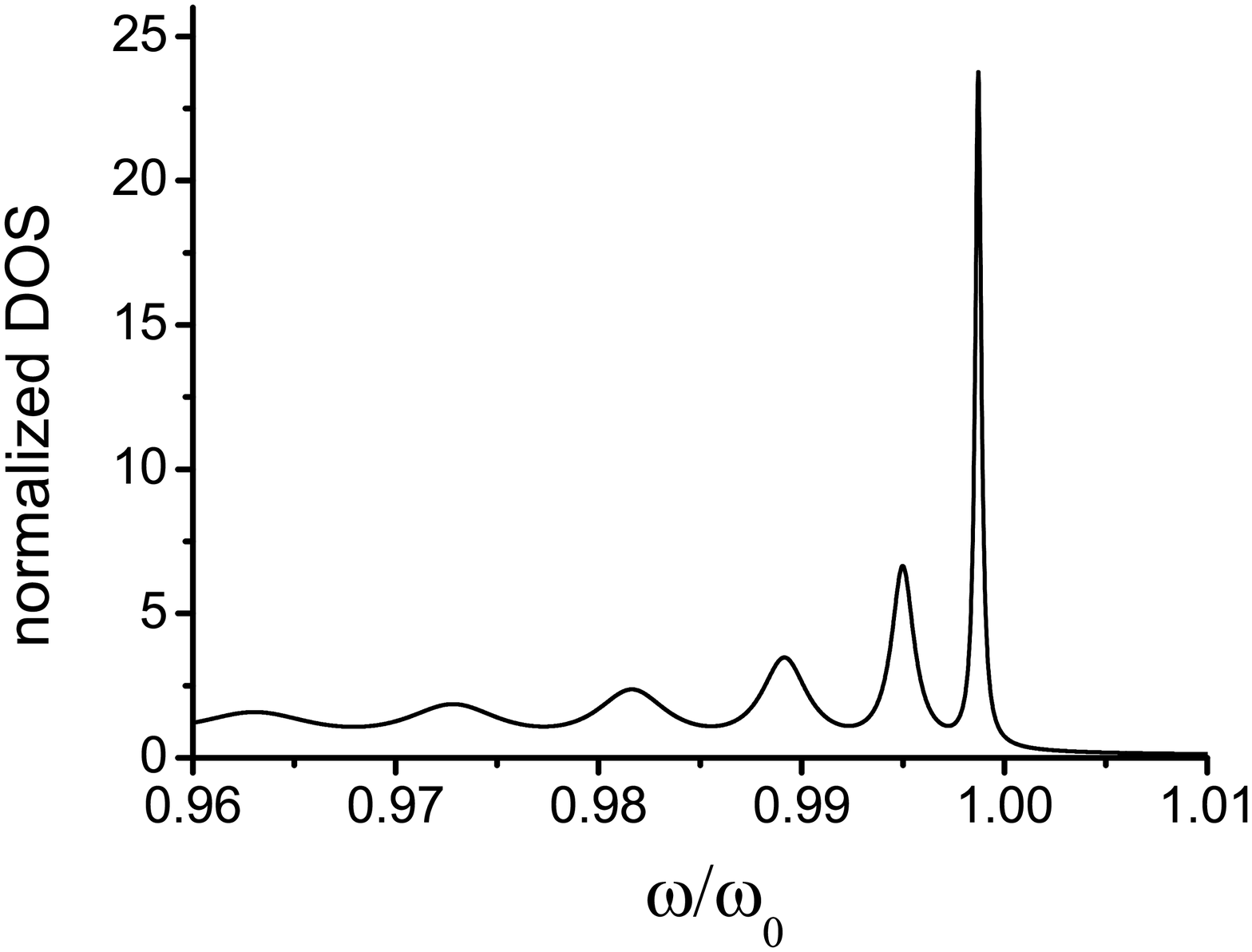}
\caption{{\bf (a)} Theoretically obtained normalized DOS using Eq. \eqref{normDOS} for $N=65$ precessions of the director and $\delta=0.091$. {\bf (b)} Theoretically obtained normalized DOS using Eq. \eqref{normDOS} for $N=40$ precessions and $\delta=0.13$.}
\label{fig:DOSth}
\end{figure}
Equations \eqref{spectrum} and \eqref{S12} allow the calculation of the emission spectrum, subject to the the determination of the DOS in the resonating structure.

First of all, Fig. \ref{fig:DOSth} depicts the DOS profile for a chiral nematic LC layer with different values of thickness and relative dielectric anisotropy, for the low frequency edge. It is shown that the value of the DOS increases with increasing product of thickness and relative dielectric anisotropy. Both these quantities determine distributed feedback within the structure. The DOS exhibits a distinct peak at the short frequency edge alongside some minor resonance peaks decreasing in magnitude with decreasing frequency. Similar results have been also reported elsewhere \cite{DOSfl}. The DOS value for the dominant edge mode diverges to infinity for a given relative dielectric anisotropy and $N\rightarrow\infty$.
\begin{figure}
\includegraphics[width=2.8in]{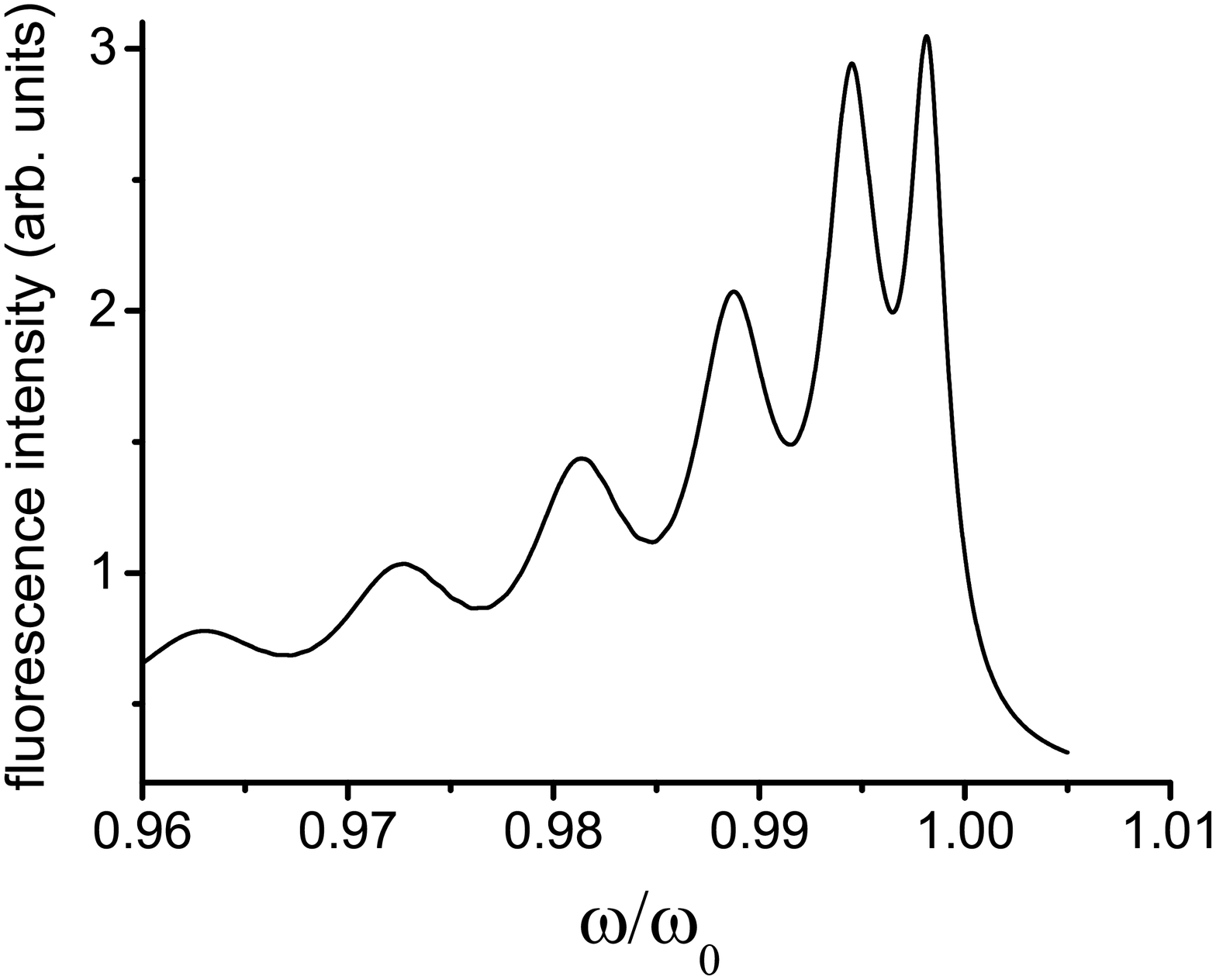}
\includegraphics[width=2.8in]{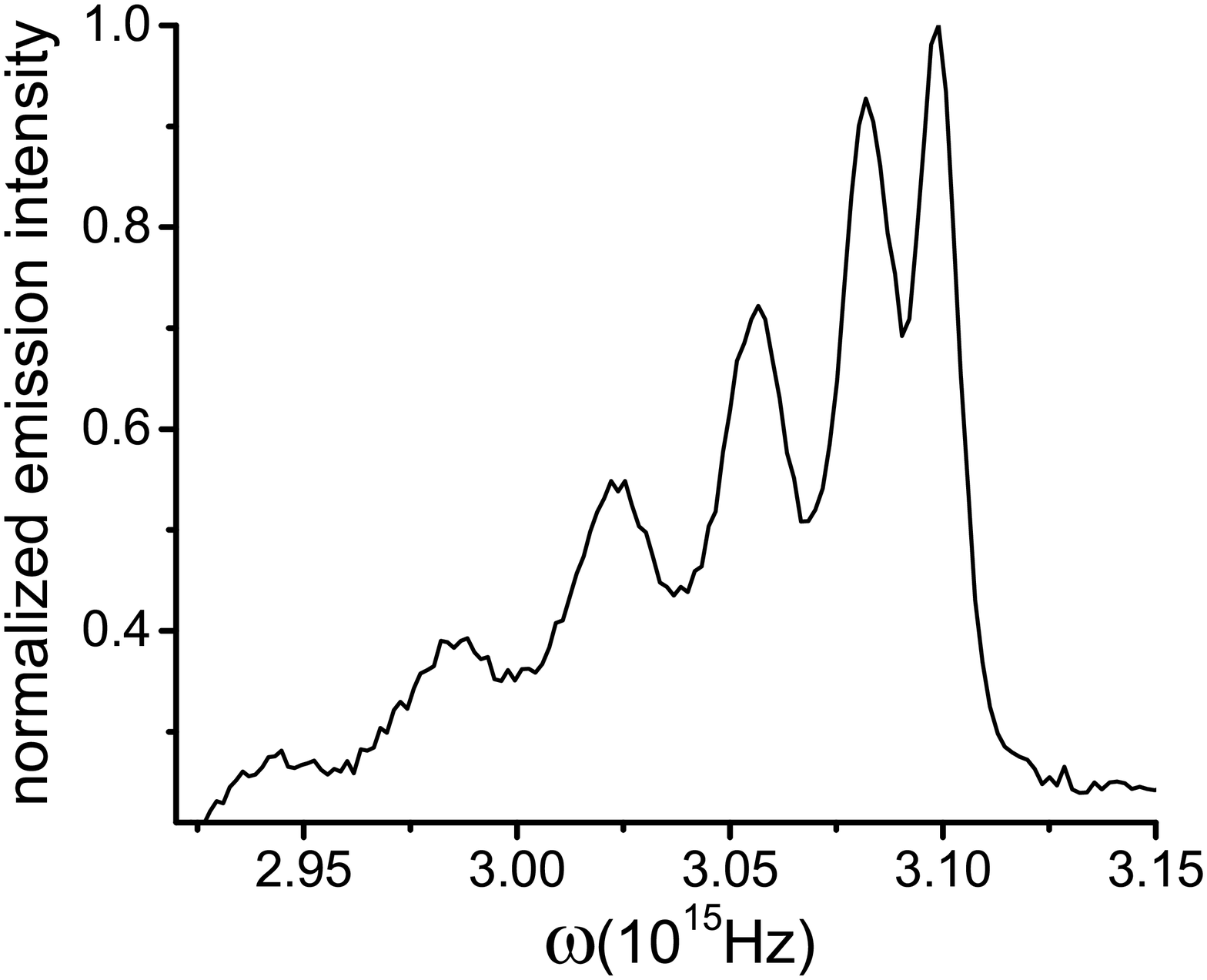}
\caption{{\bf(a)} Theoretically obtained emission spectrum for a chiral nematic LC with a gain medium for $N=40$ precessions of the director, $\delta=0.13$, $\gamma_{10}=1.25 \cdot10^{-4}\cdot\omega_0$ and $\gamma=9.36\cdot 10^{-4}\cdot\omega_0$. In this case $\omega_{10}^{\prime}=1.06\cdot\omega_0$. {\bf (b)}  Experimentally determined fluorescence spectrum obtained from a chiral nematic LC sample doped with DCM (with a fluorescence peak in the LC host at $\lambda_{\rm max}\simeq 580\,$nm) and $N\approx40$ full precessions of the molecular director. For (a) the integration limits in Eqs. \eqref{S12} are $\omega_{\rm min}=0.96\cdot\omega_0$ and $\omega_{\rm max}=1.01\cdot\omega_0$. Here, $\beta=1.05\cdot 10^{29}[{\rm SI}]$.\footnote{In all the figures that follow, we use the symbol $\omega$ instead of $\Omega$ that appears in the relations of sections \ref{sec:JCresenv} and \ref{sec:logdiv}, for convenience.} }
\label{fig:spectracomp}
\end{figure}
The DOS calculated for a sample with $N=40$ precessions of the molecular director and relative dielectric anisotropy $\delta=0.13$ is now employed to calculate the emission profile from Eq. \eqref{spectrum}. 

Figure \ref{fig:spectracomp} shows a comparison between the theoretically obtained fluorescence spectrum and that measured experimentally for a cell with thickness $L\simeq 12\;\mathrm{\mu m}$ ($N \simeq 40$) consisting of the chiral nematic LC mixture E49 doped with the fluorescent dye DCM. There is a good agreement between theoretical and experimental results, from which we deduce that the spontaneous emission spectrum exhibits significant differences from the pattern dictated by the DOS presented in Fig. \ref{fig:DOSth}. However, by accounting for the emission spectrum using the approach discussed herein appears to provide a better match with experimental observations. Our results demonstrate that we ought to account additionally for the relative position of the transition frequency with respect to the edge mode location in order to describe more accurately the emission spectrum from these periodic structures. At large oscillator strengths of the atomic transition, there is a Fano resonant mechanism between the discrete spectrum of the atomic transition and the continuum of photon states in the chiral nematic feedback structure, occurring when the atomic transition frequency lies in the region of the continuum. The same mechanism is associated with the splitting of the fluorescence into two components in the region of the logarithmic singularity due to the saddle point in the dispersion curve of a 2D photonic crystal \cite{PerlinDOS}.
\begin{figure}
\includegraphics[width=2.8in]{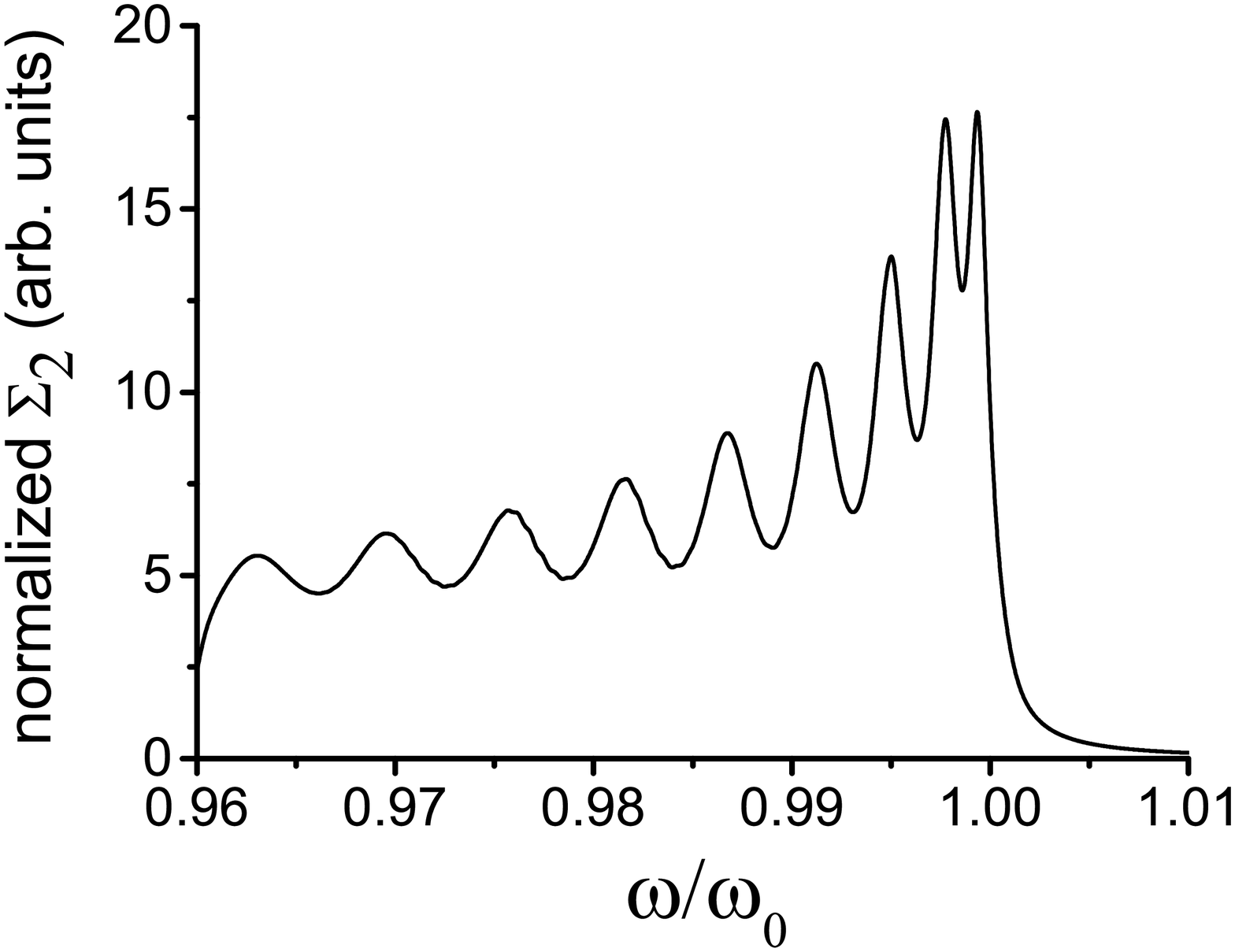}
\includegraphics[width=2.8in]{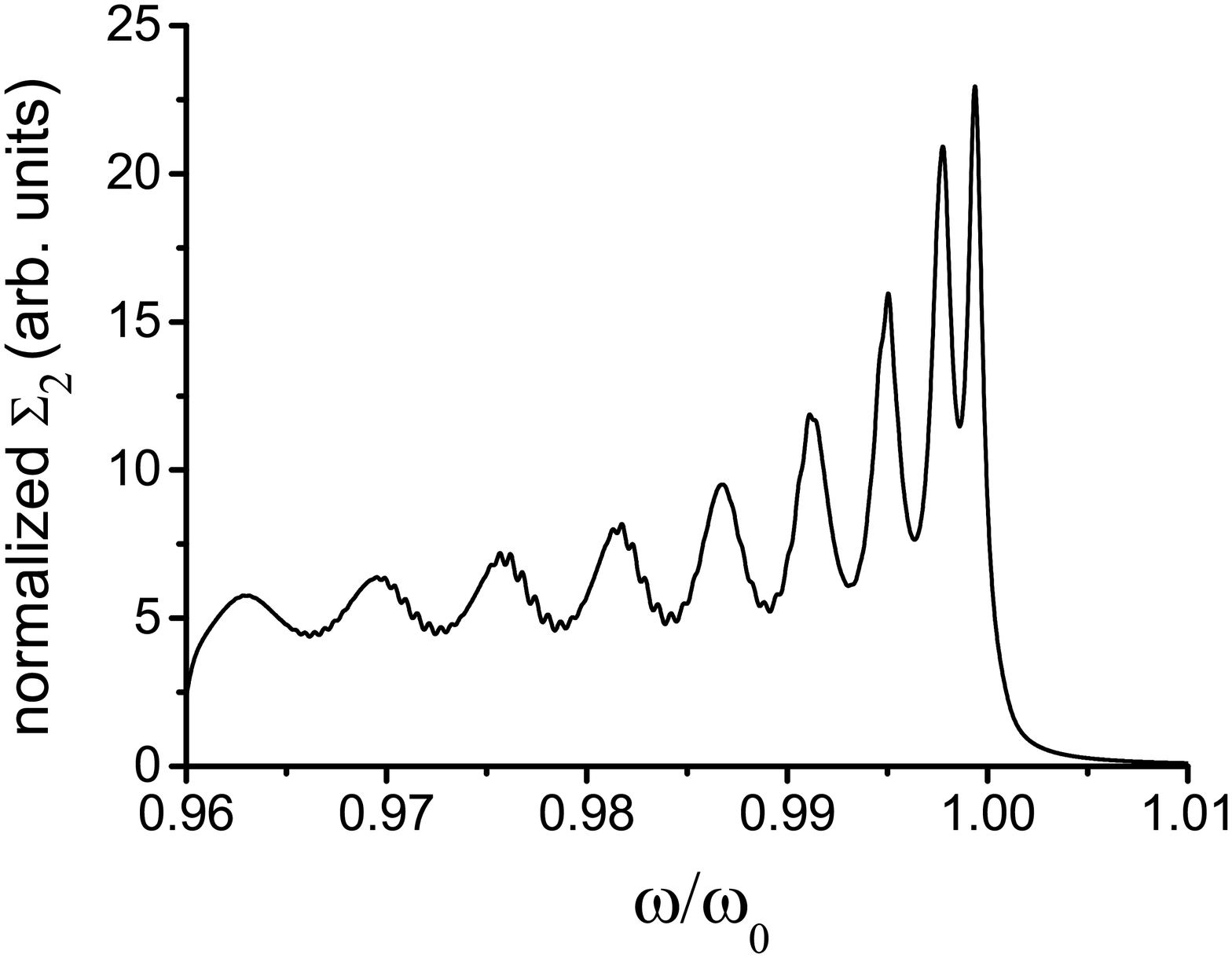}
\caption{Transition broadening $\Sigma_2$ term as a function of frequency for a chiral LC with a gain medium, for $N=60$ precessions of the director, $\delta=0.13$ and two different values of cavity losses. In {\bf (a)} $\gamma=5.55\cdot 10^{-4}\cdot\omega_0$ and in {\bf (b)} $\gamma=3.70\cdot 10^{-4}\cdot\omega_0$.}
\label{fig:S2term}
\end{figure}

Figure \ref{fig:S2term} depicts the transition broadening term $\Sigma_2$ of Eq. \eqref{S12} normalized by the same arbitrary constant, for two different values of the cavity losses. We find that apart from the change in magnitude of the term with decreasing losses, there is also a change in the relative height of the first two edge-mode peaks. Moreover, we quantify the effect of resonance for a chiral nematic LC resonator in which the feedback properties are enhanced. 
\begin{figure}
\includegraphics[width=3.4in]{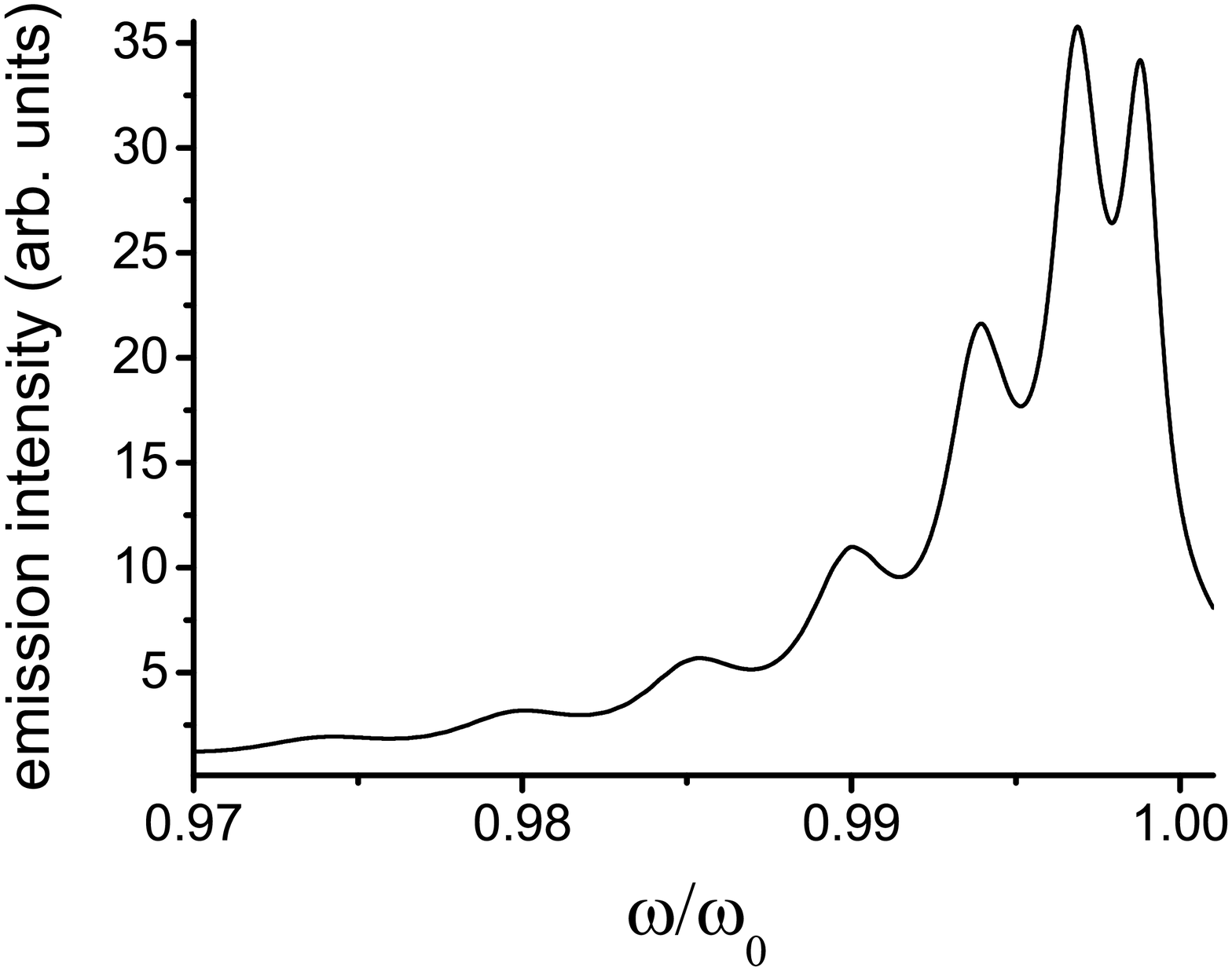}
\includegraphics[width=3.4in]{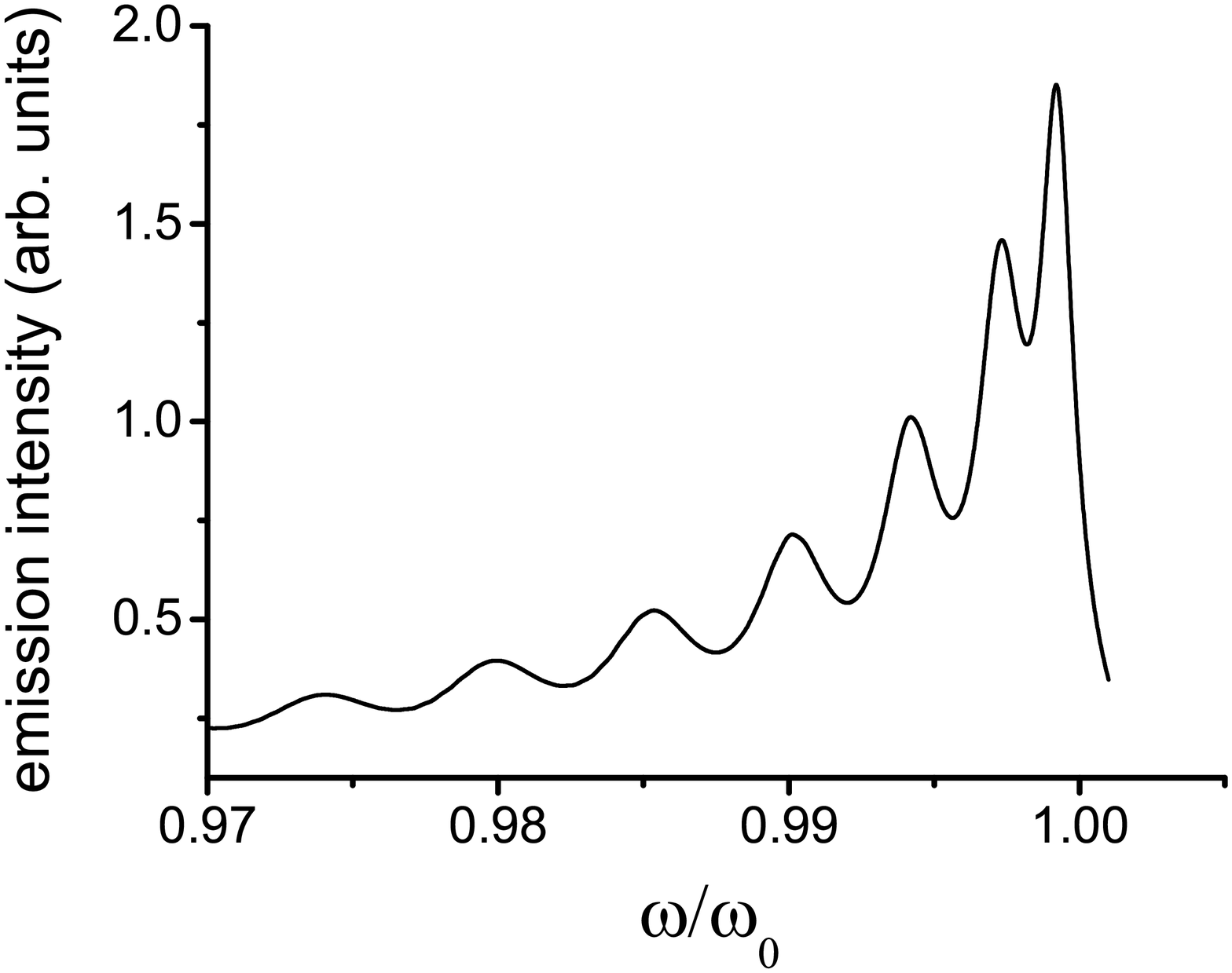}
\caption{Theoretically obtained emission spectrum for a chiral nematic LC with a gain medium for $N=65$ precessions of the director, $\delta=0.091$, $\gamma_{10}=1.25 \cdot10^{-4}\cdot\omega_0$, $\gamma=7.02\cdot10^{-4}\cdot \omega_0$ for two different detuning values. In {\bf (a)} $\omega_{10}^{\prime}=1.005\cdot \omega_0$ and in {\bf(b)} $\omega_{10}^{\prime}=1.05\cdot \omega_0$. In (a), $\omega_{\rm min}=0.95\cdot\omega_0$ and $\omega_{\rm max}=1.001\cdot\omega_0$ whereas in (b), $\omega_{\rm min}=0.97\cdot\omega_0$ and $\omega_{\rm max}=1.001\cdot\omega_0$. Here, $\beta=3 \cdot 10^{28}[{\rm SI}]$.}
\label{fig:Detspectra}
\end{figure}

Figure \ref{fig:Detspectra} depicts the emission profile calculated from Eq. \eqref{spectrum} for small and large detuning, i.e. varying the frequency offset between the electronic transition and the dominant edge mode. Our theoretical results show that the first two peaks which correspond to the two edge modes closest to the band gap are less pronounced than those in the DOS profile, when under the condition of exact resonance. This can be inferred from Fig. \ref{fig:DOSth} (a), (b) and Fig. \ref{fig:Detspectra}(a). Their magnitude also decreases with increasing detuning [Fig. \ref{fig:Detspectra}(b)]. The presence of residual attenuation due to a variety of mechanisms, such as scattering from long range thermal fluctuations of the molecular director and absorption from excited atomic levels, inhibits the feedback mechanism inside the resonator, which is manifested as a decrease in the DOS \cite{LCLasing}. This result can be also demonstrated for a Fabry-Perot resonator.

\section{Logarithmic divergence of the DOS in the region of the saddle point of a 2D photonic crystal}
\label{sec:logdiv}

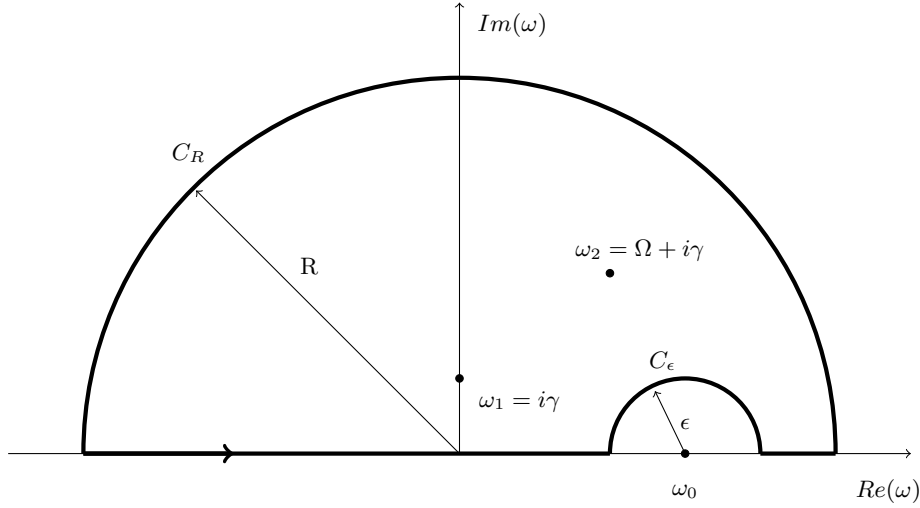
\begin{figure*}
\begin{center}
\begin{tikzpicture}
 \draw[->] (-6,0)--(6,0);
 \draw[->] (0,0)--(0,6);

 \draw[->, ultra thick] (-5,0)--(-3,0);

 \draw[->] (0,0)--(-3.5,3.5);
 \draw[->] (3,0)--(2.6,0.83);

\draw [ultra thick] (5,0)  arc[radius = 5, start angle= 0, end angle= 180];
\draw [ultra thick] (-5,0) to (2,0);
\draw [ultra thick] (4,0)  arc[radius = 1, start angle= 0, end angle= 180];
\draw [ultra thick] (4,0) to (5,0);

\node at (-2,2.5) {R};
\node at (3,0.4) {$\epsilon$};
\node[draw,circle,inner sep=1pt,fill] at (3,0){};
\node at (3,-0.5) {$\omega_0$};
\node[draw,circle,inner sep=1pt,fill] at (0,1){};
\node at (0.8,0.7) {$\omega_1=i\gamma$};
\node[draw,circle,inner sep=1pt,fill] at (2,2.4){};
\node at (2.4,2.7) {$\omega_2=\Omega+i\gamma$};

\node at (2.7,1.23) {$C_{\epsilon}$};

\node at (-3.6,4) {$C_R$};

\node at (5.7,-0.5) {$Re(\omega)$};

\node at (0.7,5.7) {$Im(\omega)$};
\end{tikzpicture}
\end{center}
\caption{Integration contour for the application of the residue theorem [Eq. \eqref{formulares}].}
\label{fig:contour}
\end{figure*}

\begin{figure}
\includegraphics[width=2.8in]{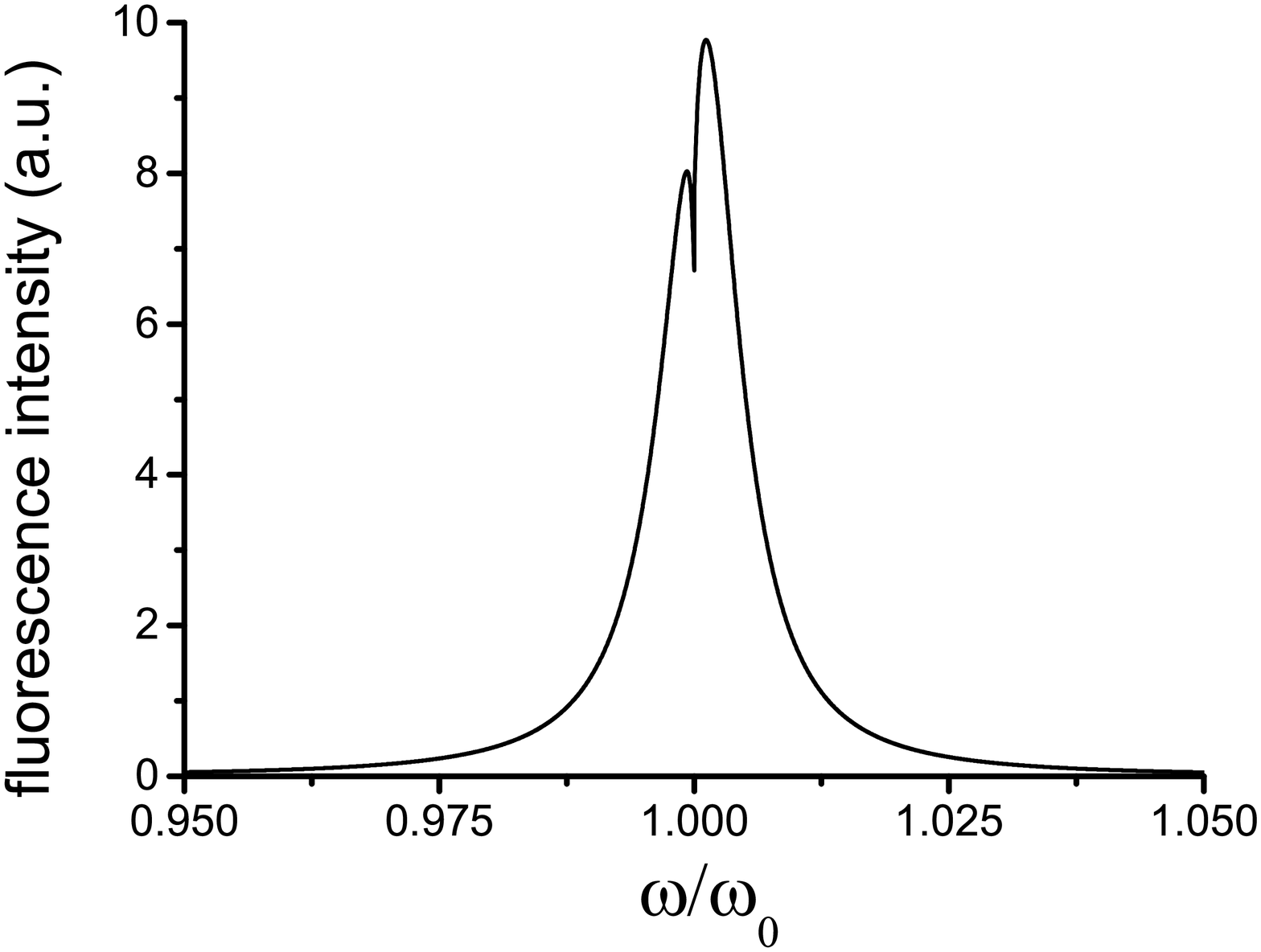}
\includegraphics[width=2.8in]{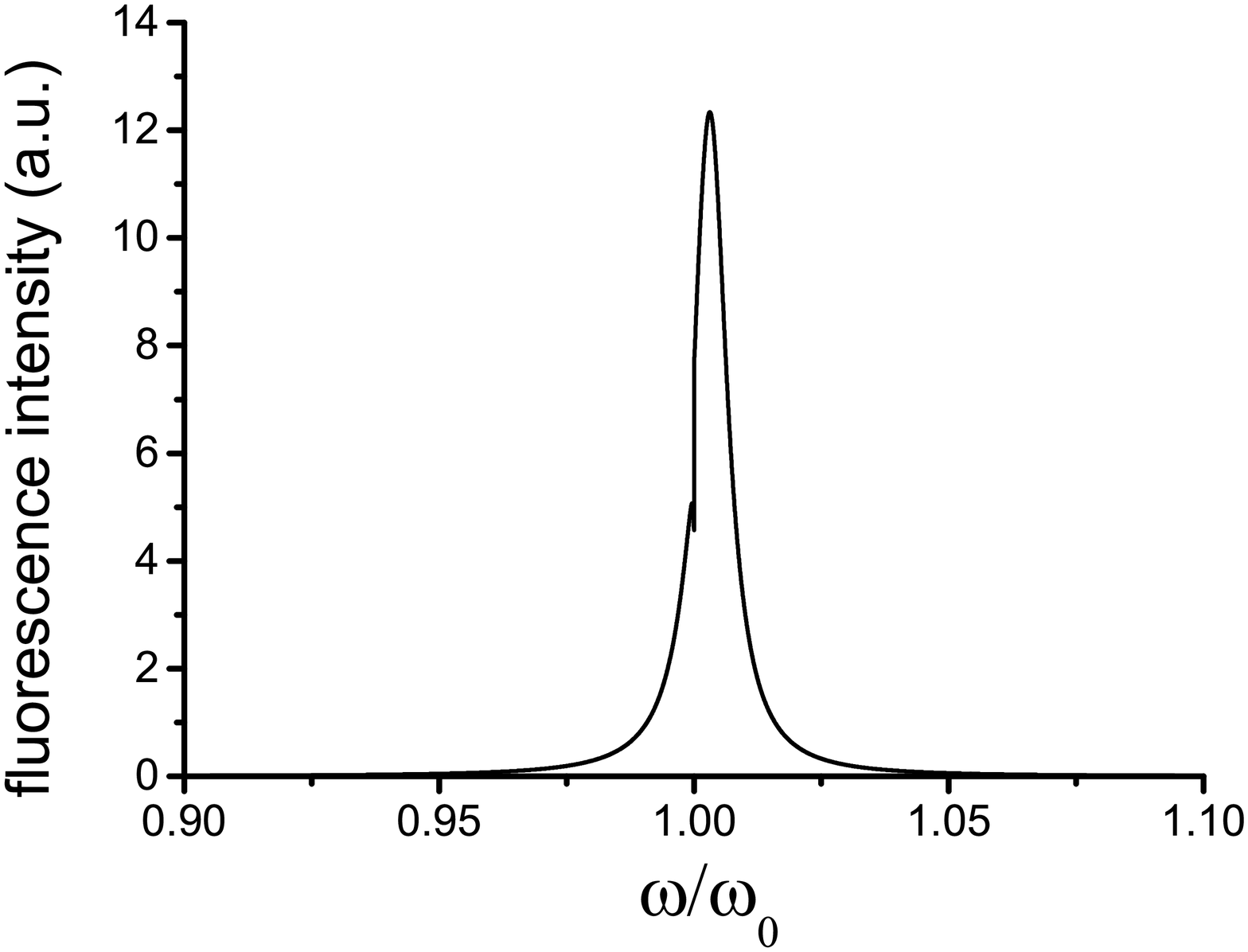}
\caption{{\bf(a)} Emission spectrum for a 2D photonic crystal with a logarithmic singularity in the DOS at $\omega_0$, as a function of frequency for  $\omega_{10}^{\prime}=1.001\cdot \omega_0$. {\bf (b)} Emission spectrum for the photonic crystal as a function of frequency for $\omega_{10}^{\prime}=1.02\cdot \omega_0$. In all cases above, $\beta^{\prime}=3 \cdot10^{27}[{\rm SI}]$, $\gamma=1.25\cdot 10^{-4}\cdot\omega_0$ and  $\gamma_{10}=1.25 \cdot10^{-5}\cdot\omega_0$.}
\label{fig:Lamb2}
\end{figure}

We will now address a particular case in which one can derive analytic expressions for the Lamb shift and the transition broadening featuring in Eq. \eqref{spectrum}. Unlike 1D photonic crystals, where the DOS displays Van Hove singularities at the band extrema, spontaneous emission in 2D crystals is not enhanced at these points despite the fact that the group velocity assumes zero values there. The exact dispersion relationship depends on the lattice of the periodic structure and the polarization of the modes considered. Here it is assumed that the atomic transition frequency is close to a saddle point ($P_1$ type) in one of the branches of the photonic band spectrum, irrespective of the lattice and the emission direction from the periodic structure. Therefore, our findings will pertain to the general case. It is known that near the saddle point in the dispersion curve of a 2D photonic crystal, the DOS exhibits a logarithmic divergence \cite{DOSPC, PerlinDOS}. We start by calculating the integral:
\begin{equation}\label{intDOS}
I=\int_{-\infty}^{+\infty}f(\omega)d\omega,\: \text{ with } \: f(\omega)=\frac{\ln\big(|a(\omega_0)(\omega-\omega_0)|\big)}{(\omega-\omega _1)(\omega-\omega _2)},
\end{equation}
where $a(\omega_0)$ has dimensions of Hz$^{-1}$ and is related to the specific photonic crystal properties and the expansion near the saddle point (with $\rho \propto -\ln[|a(\omega_0)(\omega-\omega_0)|]$) \cite{PerlinDOS}. We will find, however, that our final results are independent of this factor. Since the residue theorem is applied for single-valued functions,
we consider the branch of the logarithm defined by $\log(z)=\ln|z| + i\theta, -\frac{\pi}{2}\leq\theta<\frac{3\pi}{2}$. The integration contour is shown in Fig. \ref{fig:contour}.
According to the residue theorem, we have
\begin{align}\label{formulares}
&\int_{\omega_0+\epsilon}^R f(\omega)d\omega+\int_{C_R}f(\omega)d\omega+\int_{-R}^{\omega_0-\epsilon}f(\omega)d\omega\nonumber\\&+\int_{C_\epsilon}f(\omega)d\omega=2\pi i \left\{\frac{\log[a(\omega_2-\omega_0)]}{\omega_2-\omega_1}+\frac{\log[a(\omega_1-\omega_0)]}{\omega_1-\omega_2}\right\}\;,
\end{align}
where: $\omega_1=i\gamma$ and $\omega_2=\Omega+i\gamma$.
We also have:
\begin{align*}
&\left\lvert \int_{C_\epsilon}\frac{\log[a(\omega-\omega_0)]}{(\omega-\omega _1)(\omega-\omega _2)}d\omega\right\lvert \\
=
&\left\lvert \int_0^{\pi}\frac{\log(a\epsilon e^{i\theta})}{(\omega_0+\epsilon e^{i\theta}-\omega_1)(\omega_0+\epsilon e^{i\theta}-\omega_2)}i\epsilon e^{i\theta}d\theta\right\lvert\nonumber\\
& \leq\frac{\mid \log(a\epsilon)\mid+\pi}{(\mid\omega_0-\omega_1\mid-\epsilon)(\mid\omega_0-\omega_2\mid-\epsilon)}\pi\epsilon\rightarrow 0\;,
\end{align*}
as $\epsilon\rightarrow 0$, since $\epsilon \ln\epsilon\rightarrow 0$ when $\epsilon\rightarrow 0$.
For $R$ sufficiently large, we also have
\begin{align*}
&\left\lvert \int_{C_R}\frac{\log[a(\omega-\omega_0)]}{(\omega-\omega _1)(\omega-\omega _2)}d\omega\right\lvert\\
=
&\left\lvert \int_0^{\pi}\frac{\log[a(R e^{i\theta}-\omega_0)]}{(R e^{i\theta}-\omega_1)(R e^{i\theta}-\omega_2)}iR e^{i\theta}d\theta\right\lvert\nonumber\\
 &\leq\frac{\mid \log[a(R+\omega_0)]\mid+\pi}{(R-\mid\omega_1\mid)(R-\mid\omega_2\mid)}\pi R\rightarrow 0\;,
\end{align*}
since $\frac{\ln R}{R}\rightarrow 0$ when $R\rightarrow \infty$.
Hence, we deduce that
\begin{align*}
&\int_{-\infty}^{+\infty}f(\omega)d\omega=\int_{-\infty}^{+\infty}\frac{\ln|a(\omega-\omega_0)|}{(\omega-\omega_1)(\omega-\omega_2)}d\omega\\
&+i\pi\int_{-\infty}^{\omega_0}\frac{d\omega}{(\omega-\omega_1)(\omega-\omega_2)}\nonumber\\
&=\frac{2\pi i}{\omega_2-\omega_1}\log\left(\frac{\omega_2-\omega_0}{\omega_1-\omega_0}\right),
\end{align*} 
since
\begin{align*}
\int_{-\infty}^{\omega_0}\frac{d\omega}{(\omega-\omega_1)(\omega-\omega_2)}=\frac{1}{\omega_2-\omega_1}\log\left(\frac{\omega_2-\omega_0}{\omega_1-\omega_0}\right).
\end{align*}
We must note here that the behavior of the DOS far from the saddle point may be different (usually we assume $\rho\propto\omega$ far from the critical point). In that case, the upper integration limit is replaced by the Compton frequency \cite{PerlinDOS, BykovBook}, $\omega_c=\dfrac{mc^2}{\hbar}\cong10^{21}$ Hz, which is many orders of magnitude higher than the frequencies in the visible part of the electromagnetic spectrum. Therefore, the integration to infinity can be justified.
We conclude that:
\begin{align*}
&\int_{-\infty}^{+\infty}\frac{\ln|a(\omega-\omega_0)|}{(\omega-\omega_1)(\omega-\omega_2)}d\omega=\frac{\pi i}{\omega_2-\omega_1}\log\left(\frac{\omega_2-\omega_0}{\omega_1-\omega_0}\right)\\
&=\frac{\pi i}{\Omega}\log\left(\frac{\Omega+i\gamma-\omega_0}{i\gamma-\omega_0}\right).
\end{align*}
As we have selected the particular branch of the logarithmic function with $ -\frac{\pi}{2}\leq\theta<\frac{3\pi}{2}$, the real part of the resulting integral will have a discontinuity since the phase of the logarithm varies between $-\pi$ and $0$ for an argument selection $-\pi\leq\theta<\pi$. The same discontinuity, linked to the Lamb shift would have been exhibited if we had chosen any other branch cut outside our integration contour. The Lamb shift and the transition broadening term, in this case, read:
\begin{equation}\label{S12log}
\begin{aligned}
&\Sigma_1={\beta^{\prime}}|{\hat{\mathbf{e}}_{kl}\hat{\mathbf{r}}_{10}}|^2\Re\left\{\int_{-\infty}^{+\infty}\frac{\ln|a(\omega-\omega_0)|}{(i\gamma-\omega)(\omega-\Omega-i\gamma)}d\omega\right\},\\
&\Sigma_2={\beta^{\prime}}|{\hat{\mathbf{e}}_{kl}\hat{\mathbf{r}}_{10}}|^2\Im\left\{\int_{-\infty}^{+\infty}\frac{\ln|a(\omega-\omega_0)|}{(i\gamma-\omega)(\omega-\Omega-i\gamma)}d\omega\right\},
\end{aligned}
\end{equation}
where we have assumed that the normalization term $\beta$ in Eqs. \eqref{S12} is modified by some parameters particular to the expression of the DOS for a photonic crystal \cite{PerlinDOS}, to yield $\beta^{\prime}$. We can deduce that for 2D photonic crystals in the saddle point of the dispersion function, resonance is associated with a Lamb dip in the fluorescence spectrum and a split in the real part of the Fourier transform of the Green's function. 
\begin{figure}
\includegraphics[width=2.8in]{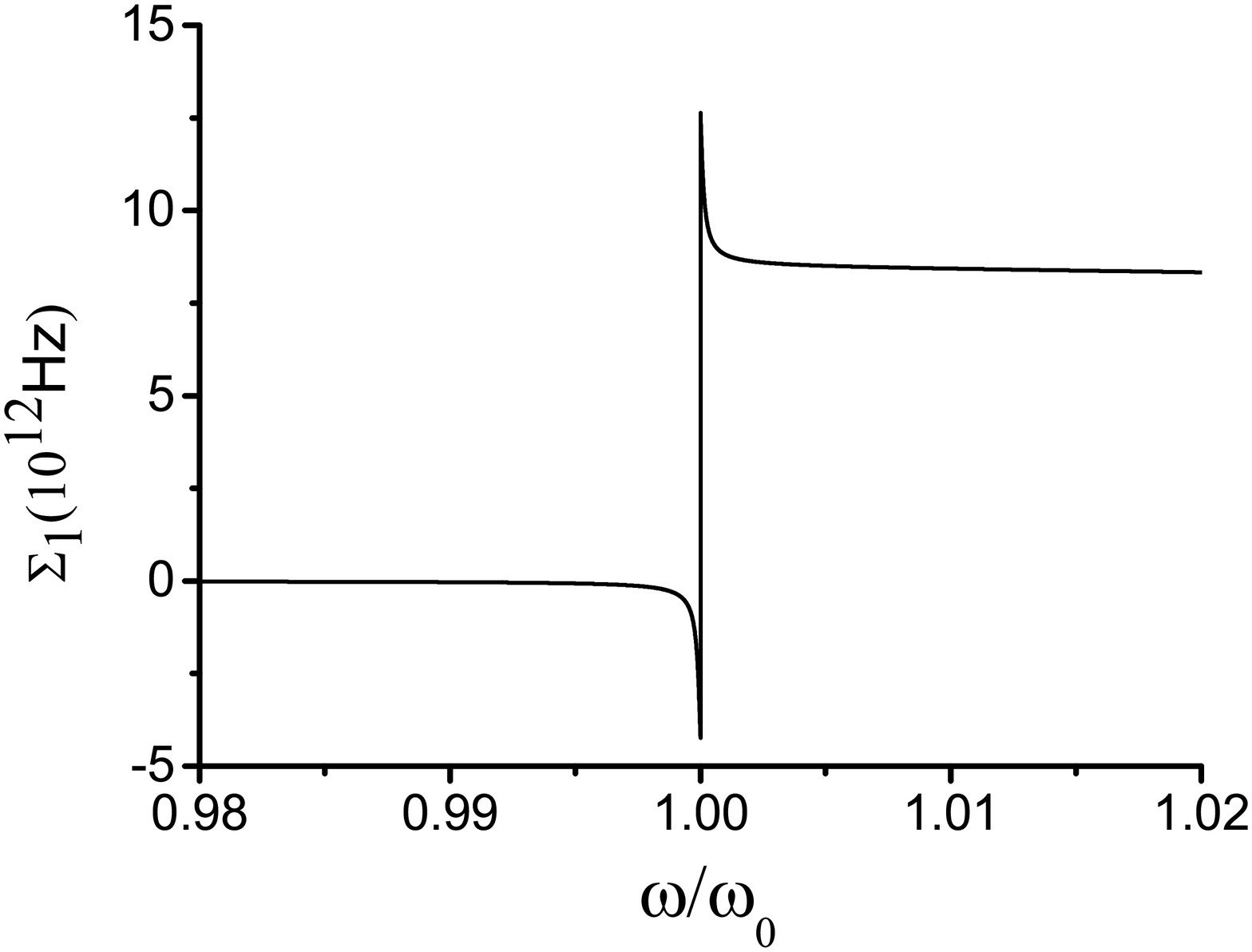}
\includegraphics[width=2.8in]{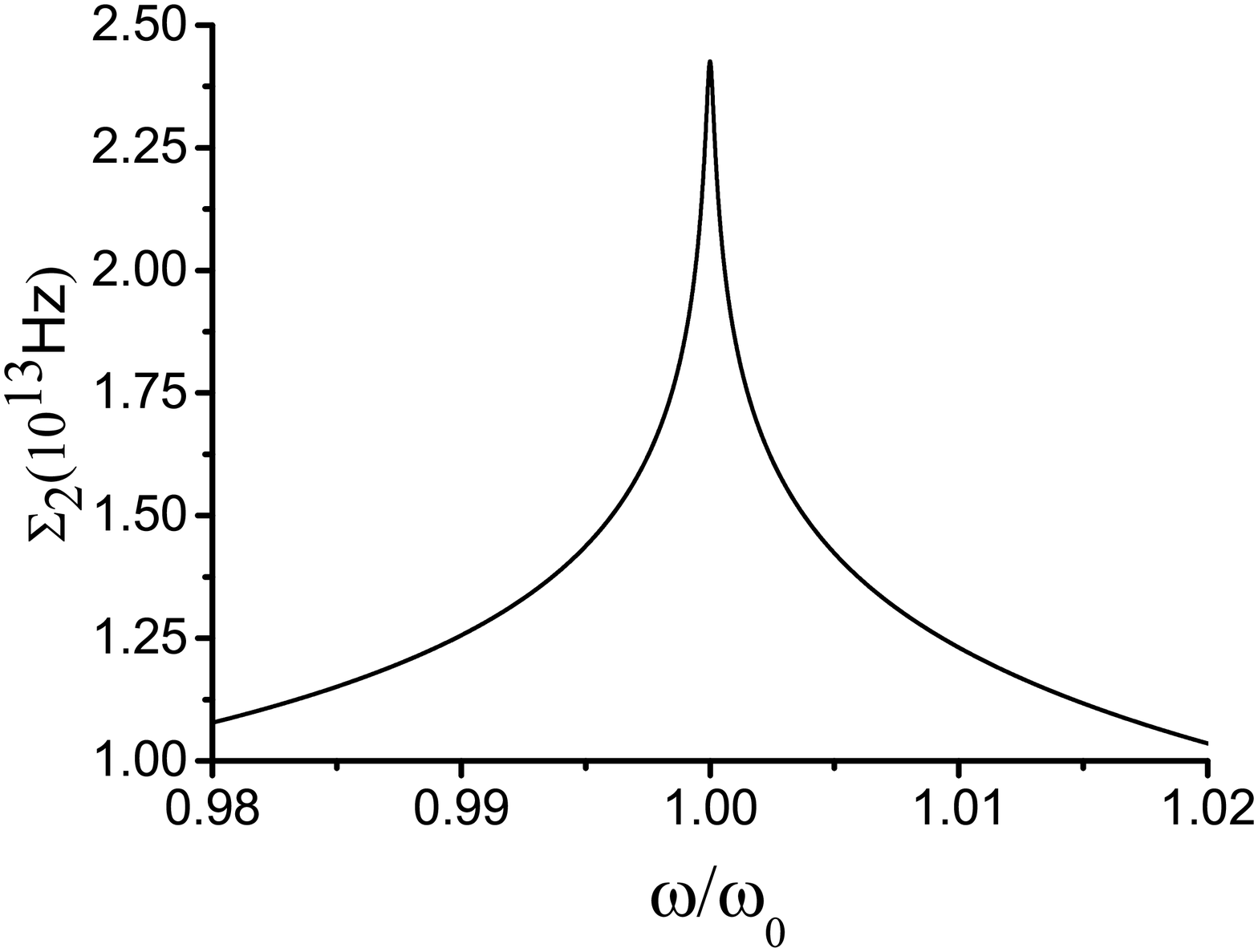}
\caption{{\bf (a)} The Lamb shift term ($\Sigma_1$) as a function of frequency for a 2D photonic crystal with a logarithmic singularity in the DOS at $\omega_0$. {\bf (b)} Transition broadening ($\Sigma_2$) as a function of frequency for a photonic crystal with a logarithmic singularity in the DOS at $\omega_0$. In all cases above, $\omega_{10}^{\prime}=1.001\cdot \omega_0$ and $\gamma=1.25\cdot 10^{-4}\cdot\omega_0$.}
\label{fig:Lamb1}
\end{figure}
The split (term $\Sigma_1$) here is due to the behaviour of the complex logarithmic function; however, as we can observe in Fig. \ref{fig:Lamb1} the magnitude of that term is significantly lower than the broadening term $\Sigma_2$ of the transition $\Ket{1,\{0\}}\rightarrow\Ket{0,\{1\}_{\kappa}}$. The Lamb dip essentially vanishes for larger detunings, as we can see in the emission spectra of Fig. \ref{fig:Lamb2}. A Lamb shift has also been reported inside the (complete) photonic band gap for hydrogenic atoms embedded in 1D periodic structures \cite{SajeevPRL}.

\section{Discussion of results}
\label{sec:discussion}

Concerning the validity of our results, we note that the rotating wave approximation we relied upon, requires a small detuning from the atomic transition frequency, in order for the `non-resonant' terms to be much smaller than the `resonant' ones when averaging over a time period of the order of $1/\omega$ in the interaction picture \cite{RWAIrish}. This constraint is met by selecting an appropriate upper and lower integration limit in Eqs. \eqref{S12} and by the presence of the function $\log|a(\omega-\omega_0)|$ in Eqs. \ref{S12log}, ensuring that the major contribution to the integral originates from the region $|\omega-\omega_{10}|\ll(\omega+\omega_{10})$. Likewise, the most significant contribution in the emission from chiral nematics will be from the first two edge mode peaks on the same side of the band gap, for $N\delta\gg1$. At this point, we ought to mention that the two-level system approach is certainly a considerable simplification for fluorescent dyes. These complex molecules have states with many vibrational and rotational levels determining their spectra. In this analysis, mechanisms such as triple state generation and resonant energy transfer have also been ignored. Under the assumption of very small detuning, the effect of these phenomena can be quantified through the introduction of an imaginary part in the mode frequency. Recently, rare-earth-doped nanocrystals have been used as the gain medium hosted in chiral nematic LCs \cite{SPIEpaper}. In this case, the assumption of a two-level atom interacting with the electromagnetic field maybe much more appropriate to describe spontaneous emission from the resonator. Employing the DOS in the JC model is then expected to lead to a better match between theory and experiment. Moreover, for a further insight to spontaneous emission from such periodic structures, we could resort to the resolvent method \cite{BykovBook}, in which the matrix elements of the time evolution operator can be calculated.

\section{Conclusions}
\label{sec:consclusions}

In this work, we investigated resonance in a partial distributed feedback structure, for the diffractive polarization. We found that there is a disparity between the DOS and the calculated emission spectrum, which is also verified experimentally. We outlined the main effects occurring for different detuning values through deriving analytic results for the Lamb shift and the transition broadening following the logarithmic divergence of the DOS in a 2D photonic crystal. We conclude that incorporating cavity losses alongside broadening mechanisms, and including the effect of resonance leads to a more comprehensive treatment of spontaneous emission from these structures.

\begin{acknowledgments}
The authors gratefully acknowledge the Engineering and Physical Sciences Research Council (UK) for financial support
through the Photonics Systems Development Centre for Doctoral Training. One of the authors (S.M.M.) gratefully acknowledges The Royal Society for financial support. One of the authors (T.K.M.) gratefully acknowledges the Onassis Foundation for financial support.
\end{acknowledgments}

\appendix\section{Derivation of the equations of motion for the probability amplitudes}
\label{sec:AppA}
\setcounter{equation}{0}
\numberwithin{equation}{section}

Here we will derive the equations of motion for the time varying amplitudes $c_1(t)$ and $c^{\kappa}_{0}(t)$. For the left-hand side of Schr\"{o}dinger's equation applied for the state $\Ket\psi$ in Eq. \eqref{wfunc}, we have:
\begin{equation}\label{Schreq}
\begin{aligned}
  H\Ket\psi&=\exp(-i\omega_{10}t/2)\bigg[\frac{1}{2}\hbar\omega_{10}c_1(\mathbf{R},t)\Ket{1,\{0\}}\\
  &-\frac{1}{2}\hbar\omega_{10}\sum_{\kappa}\exp(i\delta_{\kappa}t)c^{\kappa}_{0}(\mathbf{R},t)\Ket{0,\{1\}_{\kappa}}\\
  &+  \hbar\sum_{\kappa}\omega_{\kappa}\exp(i\delta_{\kappa}t)c^{\kappa}_{0}(\mathbf{R},t)\Ket{0,\{1\}_{\kappa}} \\
  &+i\sum_{\kappa}\mu_{\kappa}c_{1}(\mathbf{R},t)\Ket{0,\{1\}_{\kappa}} \\
  &-i\sum_{\kappa}\mu_{\kappa}^{*}\exp(i\delta_{\kappa}t)c^{\kappa}_{0}(\mathbf{R},t)\Ket{1,\{0\}}\bigg].
  \end{aligned}
\end{equation}
The right hand side of Schr\"{o}dinger's equation \eqref{Schreq} reads:
\begin{equation}\label{RHS}
\begin{aligned}
  i\hbar\frac{\partial\psi}{\partial t}&=i\hbar\cdot \exp(-i\omega_{10}t/2)\frac{dc_1}{dt}(\mathbf{R},t)\Ket{1,\{0\}}\\
 & +\frac{1}{2}\hbar\omega_{10}\cdot \exp(-i\omega_{10}t/2)c_1(\mathbf{R},t)\Ket{1,\{0\}}\nonumber\\
&+ \exp(-i\omega_{10}t/2)\bigg[\sum_{\kappa}i\hbar\cdot \exp(i\delta_{\kappa}t)\frac{dc^{\kappa}_{0}}{dt}(\mathbf{R},t)\Ket{0,\{1\}_{\kappa}}\\
&-\hbar(\omega_{10}-\omega_{\kappa})\exp(i\delta_{\kappa}t)c^{\kappa}_{0}(\mathbf{R},t)\Ket{0,\{1\}_{\kappa}}\bigg] \\
& +\frac{1}{2}\hbar\omega_{10}\exp(-i\omega_{10}t/2)\sum_{\kappa}\exp(i\delta_{\kappa}t)c^{\kappa}_{0}(\mathbf{R},t)\Ket{0,\{1\}_{\kappa}}\\
& =i\hbar\cdot  \exp(-i\omega_{10}t/2)\frac{dc_1}{dt}(\mathbf{R},t)\Ket{1,\{0\}}\\
 &+\frac{1}{2}\hbar\omega_{10}\exp(-i\omega_{10}t/2)c_1(\mathbf{R},t)\Ket{1,\{0\}}\\
  &+ \exp(-i\omega_{10}t/2)\bigg[\sum_{\kappa}i\hbar\cdot \exp(i\delta_{\kappa}t)\frac{dc^{\kappa}_{0}}{dt}(\mathbf{R},t)\Ket{0,\{1\}_{\kappa}}\\
  &+\hbar\omega_{\kappa}\exp(i\delta_{\kappa}t)c^{\kappa}_{0}(\mathbf{R},t)\Ket{0,\{1\}_{\kappa}}\bigg]\\
 &-\frac{1}{2}\hbar\omega_{10}\cdot \exp(-i\omega_{10}t/2)\sum_{\kappa}\exp(i\delta_{\kappa}t)c^{\kappa}_{0}(\mathbf{R},t)\Ket{0,\{1\}_{\kappa}}.
\end{aligned}
\end{equation}
Upon canceling the common terms, one obtains:
\begin{align}\label{Step1}
   i\sum_{\kappa}\mu_{\kappa}c_{1}(\mathbf{R},t)\Ket{0,\{1\}_{\kappa}}
  -i\sum_{\kappa}\mu_{\kappa}^{*}\exp(i\delta_{\kappa}t)c^{\kappa}_{0}(\mathbf{R},t)\Ket{1,\{0\}}\nonumber\\
  =i\hbar\frac{dc_1}{dt}(\mathbf{R},t)\Ket{1,\{0\}}+\sum_{\kappa}i\hbar\cdot\exp(i\delta_{\kappa}t)\frac{dc^{\kappa}_{0}}{dt}(\mathbf{R},t)\Ket{0,\{1\}_{\kappa}}.
\end{align}
After taking the inner product with the states $\big(\Ket{0,\{1\}_{\kappa}}\big)^{\dagger}$ and $\big(\Ket{1,\{0\}}\big)^{\dagger}$ we arrive at the equations of motion presented in \cite{PerlinDOS}:
\begin{equation}\label{EOM}
  \begin{cases}\dfrac{dc_1}{dt}(\mathbf{R},t)=-\sum_{\kappa}\mu_{\kappa}^{*}\exp(i\delta_{\kappa}t)c^{\kappa}_{0}(\mathbf{R},t), \vspace{3mm} \\ 
  \dfrac{dc^{\kappa}_{0}}{dt}(\mathbf{R},t)= \exp(-i\delta_{\kappa}t)\mu_{\kappa}c_{1}(\mathbf{R},t).\end{cases}
  \end{equation}

\section{Derivation of the emission spectrum}
\label{sec:AppB}
\setcounter{equation}{0}
\numberwithin{equation}{section}

In this section we will derive an expression for the Fourier transform of the Green's function, used in Eq. \eqref{spectrum} for the determination of the emission spectrum. Applying the Fourier transform to both sides of Eq. \eqref{dercoeff} and invoking the convolution theorem, we obtain [for $c_1(t=0)=1$]:
\begin{equation*}
\begin{aligned}
&\mathcal{F}\left\{\frac{dc_1}{dt}\right\}(\mathbf{R},t)=-\mathcal{F}\{g(\mathbf{R},t)\}\mathcal{F}\{c_1(\mathbf{R},t)\},\\
\\
& -i(\Omega-\omega_{10})\tilde{c_1}(\Omega-\omega_{10})-c_1(t=0)\\
&=-\tilde{g}(\mathbf{R},\Omega-\omega_{10})\tilde{c_1}(\Omega-\omega_{10}),\\
\\
& -i(\Omega-\acute{\omega_{10}}+i\gamma_{10})\tilde{c_1}(\Omega-\omega_{10})-1\\
& =-\tilde{g}(\mathbf{R},\Omega-\omega_{10})\tilde{c_1}(\Omega-\omega_{10}).
\end{aligned}
\end{equation*}
Hence:
\begin{align}
\tilde{c_1}(\Omega-\omega_{10})=\frac{1}{\gamma_{10}-i(\Omega-\acute{\omega_{10}})+\tilde{g}(\mathbf{R},\Omega-\omega_{10})}\;,
\end{align}
where:
\begin{equation}
\tilde{f}(\omega)=\int_0^{\infty}f(t)e^{i\omega t}dt.
\end{equation}
We also have:
\begin{equation}
\mathcal{F}\{g(\mathbf{R},t)\}=\beta \int_{\omega_{\rm min}}^{\omega_{\rm max}}\frac{\mathcal{F}\{u(t)\exp{(At)}\}}{\omega-i\gamma}\rho_{l}(\mathbf{R},\omega)d\omega,
\end{equation}
hence:
\begin{equation}\label{intform}
\tilde{g}(\mathbf{R},\Omega-\omega_{10})=-\beta\int_{\omega_{\rm min}}^{\omega_{\rm max}}\frac{\rho_{l}(\mathbf{R},\omega)}{D}d\omega,
\end{equation}
where: 
\begin{align*}
&D=(\omega-i\gamma)\big[(\Re\{A\}-\Im\{\Omega-\omega_{10}\})\\
&+i(\Im\{A\}+\Re\{\Omega-\omega_{10}\})\big],
\end{align*}
and $A=\gamma_{10}-\gamma+i (\acute{\omega_{10}}-\omega)$.\\

\noindent The integral \eqref{intform} above can be recast in the form:
\begin{equation}
\tilde{g}(\mathbf{R},\Omega-\omega_{10})=-i\beta\int_{\omega_{\rm min}}^{\omega_{\rm max}}\frac{\rho_{l}(\mathbf{R},\omega)}{(\omega-i\gamma)(\omega-\Omega-i\gamma)}d\omega,
\end{equation}
which is used in the main text.

\vspace{5mm}

\begin{center}
{\bf *****}
\end{center}
\end{document}